\documentclass[aps,prd,twocolumn,showpacs,amsmath,amssymb]{revtex4}

\begin{document}

\title{Extension of warm inflation to noncanonical scalar fields}

\author{Xiao-Min Zhang}
\email{zhangxm@mail.bnu.edu.cn}
\affiliation{Department of Physics, Beijing Normal University, Beijing 100875, China}
\author{Jian-Yang Zhu}
\thanks{Corresponding author}
\email{zhujy@bnu.edu.cn}
\affiliation{Department of Physics, Beijing Normal University, Beijing 100875, China}
\date{\today}

\begin{abstract}
We extend the warm inflationary scenario to the case of the noncanonical scalar fields. The equation of
motion and the other basic equations of this new scenario are obtained. The Hubble damped term is enhanced in
noncanonical inflation. A linear stability analysis is performed to give the proper slow-roll conditions in warm
noncanonical inflation. We study the density fluctuations in the new picture and obtain an approximate analytic
expression of the power spectrum. The energy scale at the horizon crossing is depressed by both noncanonical
effect and thermal effect, and so is the tensor-to-scalar ratio. Besides the synergy, the noncanonical effect and
the thermal effect are competing in the case of the warm noncanonical inflation.
\end{abstract}
\maketitle

\section{Introduction}

Inflation is a quasiexponential expansion (strictly speaking, the inflation is
an accelerated expansion and often taken to be a regime of quasiexponential
expansion in the majority of models considered in the literatures) in the
very early Universe \cite{Guth1981,Linde1982,Albrecht1982}, which can give
successful explanation to the problems such as the horizon and the flatness. As a
necessary supplement to the standard cosmological model, the inflation can also
produce seeds to give rise to the large scale structure and to the observed
little anisotropy of cosmological microwave background (CMB) \cite
{WMAP,PLANCK} through vacuum fluctuations. Besides the standard inflation,
there is also another type of inflation called warm inflation which is
proposed by Berera and Fang \cite{BereraFang}. Radiation is produced
constantly through the interaction ${\cal L}_{int}$ between the inflaton
field and other subdominated boson or fermion fields during warm inflation
so there is no reheating phase. The Universe can smoothly go into the big-bang
phase. And the density fluctuations originate mainly from the thermal
fluctuations \cite{BereraFang,Lisa2004,Berera2000} rather than vacuum
fluctuation. Many problems suffered in standard inflation such as eta
problem \cite{etaproblem,BereraIanRamos} and the overlarge amplitude of the
inflaton \cite{Berera2005,BereraIanRamos} can be cured in warm inflation.
With an additional thermal damped term $\Gamma\dot\phi$ added to the
evolution equation of the inflaton, the slow-roll conditions are much more
easily satisfied \cite{Ian2008,Campo2010,ZhangZhu}.

Usually, the inflation can be realized using the canonical scalar field which has the
Lagrangian density ${\cal L}=X-V_0$, where $X=\frac12 g^{\mu\nu}\partial_{%
\mu}\phi\partial_{\nu}\phi$ and $V_0$ is the potential of inflaton. But
the noncanonical fields have many novel features as the inflaton when the
Universe accelerates, such as that the equations of motion remain second
order and that the slow-roll conditions become more easily satisfied
compared to canonical inflationary theory \cite{refining2012}. The
tensor-to-scalar ratio can drop considerably in most plausible noncanonical
models \cite{refining2012,Cai2011} or increase in some phenomenological
models \cite{Mukhanov2006}. Much work has been done about noncanonical
standard inflation \cite
{refining2012,Mukhanov2006,Armendariz-Picon,Garriga1999,Gwyn2013,Tzirakis,Franche2010,Eassona2013,Bean2008}%
, and noncanonical fields are the more universal case with a general
Lagrangian density satisfied some conditions \cite{Franche2010}. However,
warm inflation as a kind of new and realizable inflationary scenario, has always dealt with canonical fields except in \cite{Cai2011} where a warm
Dirac-Born-Infeld (DBI) inflationary model was proposed. In this paper we
try to extend warm inflation to a general noncanonical scalar field and
thus the inflation can have a greater and broader scope. Through the new
picture, we can find whether its predictions can be fitted to the
observation and what attractive and new features can be obtained.

The paper is organized as follows: In Sec. \ref{sec2}, we introduce a new
noncanonical warm inflationary scenario and get the basic equations of the
new picture. Then in Sec. \ref{sec3}, we propose slow-roll inflation in the new picture and make a
fully linear stability analysis to obtain the conditions that guarantee the
slow-roll approximation is valid. The scalar and tensor
perturbations in the new scenario are performed in Sec. \ref{sec4}. Finally,
we draw the conclusions in Sec. \ref{sec5}.

\section{noncanonical warm inflationary scenario}\label{sec2}

In warm inflationary case, the Universe is a multicomponent
system; thus, the total matter action can be given as:
\begin{equation}
S=\int d^4x\sqrt{-g}\left[ {\cal L}(X,\phi )+{\cal L}_R+{\cal L}%
_{int}\right] ,  \label{action}
\end{equation}
where the Lagrangian density of the noncanonical field is ${\cal L}%
_{non-can}={\cal L}(X,\phi )$, which can be an arbitrary function of the
inflaton field $\phi $ and the kinetic term $X$, and for brevity we use $%
{\cal L}$ to stand for ${\cal L}(X,\phi )$, ${\cal L}_R$ denotes the
Lagrangian of the radiation fields and ${\cal L}_{int}$ denotes the
interaction term between inflaton and other fields. In order to have a
uniform normalization of the field, we will make the Lagrangian density in a
form that can reduce to canonical case (i.e., ${\cal L}=X-V_0$) in small $X$
limit. The noncanonical Lagrangian density should satisfy the conditions: $%
{\cal L}_X\geq 0$ and ${\cal L}_{XX}\geq 0$ (where a subscript $X$ here
denotes a derivative while the subscripts in ${\cal L}_R$ and ${\cal L}_{int}
$ are just labels) to obey the null energy condition and the physical propagation of
perturbations \cite{Franche2010,Bean2008}. Through these two conditions and
normalization of the field, we can obtain ${\cal L}_X\geq 1$. The equation
of motion can be obtained by taking the variation of the action:
\begin{equation}
\frac{\partial \left( {\cal L}(X,\phi )+{\cal L}_{int}\right) }{\partial
\phi }-\frac 1{\sqrt{-g}}\partial _\mu \left[ \sqrt{-g}\frac{\partial {\cal L%
}(X,\phi )}{\partial (\partial _\mu \phi )}\right] =0.  \label{EOM0}
\end{equation}
In the spatially flat Friedmann-Robertson-Walker Universe, the mean
inflaton field is homogeneous, i.e. $\phi =\phi (t)$; hence the equation of
motion reduces to
\begin{eqnarray}
\left[ \frac{\partial {\cal L}(X,\phi )}{\partial X}+2X\frac{\partial ^2%
{\cal L}(X,\phi )}{\partial X^2}\right] \ddot{\phi} &&  \nonumber
\label{EOM1} \\
+\left[ 3H\frac{\partial {\cal L}(X,\phi )}{\partial X}+\dot{\phi}\frac{%
\partial ^2{\cal L}(X,\phi )}{\partial X\partial \phi }\right] \dot{\phi} &&
\nonumber \\
-\frac{\partial \left( {\cal L}(X,\phi )+{\cal L}_{int}\right) }{\partial
\phi } &=&0,
\end{eqnarray}
where $X=\frac 12\dot{\phi}^2$. Through the energy-momentum tensor of $\phi $
$:T^{\mu \nu }=\left( \partial {\cal L}/\partial X\right) \left( \partial
^\mu \phi \partial ^\nu \phi \right) -g^{\mu \nu }{\cal L}$, we can get the
energy density and pressure of the field: $\rho (\phi ,X)=2X\left( \partial
{\cal L}/\partial X\right) -{\cal L}$, $p(\phi ,X)={\cal L}$. An important
parameter of the noncanonical field is the sound speed which can describe the
traveling speed of scalar perturbations: $c_s^2=p_X(\phi ,X)/\rho _X(\phi
,X)=\left( 1+2X{\cal L}_{XX}/{\cal L}_X\right) ^{-1}$, where the subscript $X
$ denotes a derivative.

Now we consider a special case that the Lagrangian density can be written in a separable form
for the kinetic term and the potential term, i.e. ${\cal L}=K(X)-V_0(\phi
)$, where $K$ is the noncanonical kinetic term that is weakly dependent or
independent on $\phi $ \cite{Franche2010}, so we assume $K$ is only the
function of $X$. In this case we have ${\cal L}_{X\phi }=0$ and $K_X={\cal L}%
_X$. The general Lagrangian mainly contains two kinds: a series-form Lagrangian
and a closed-form Lagrangian \cite{Franche2010}.The second form can be reduced to
a canonical or DBI inflation in the specific gauge ${\cal L}_X=c_s^{-1}$ \cite
{Bean2008}. The interaction term ${\cal L}_{int}$ in Eq. (\ref{action}) is
only the function of zero order of the inflaton and other fields but not of
the derivative of the fields. The most successful explanation of the
interaction between the inflaton and other fields is the supersymmetric
two-stage mechanism \cite{BereraKephart,MossXiong}. We use $\Gamma \dot{\phi}
$ to describe the dissipation effect of $\phi $ to all other fields \cite
{BereraFang,Berera2005,Berera2000,BereraIanRamos}, which is a thermal
damping term. The other terms that do not contain $\dot{\phi}$ in the $%
\partial {\cal L}_{int}/\partial \phi $ of Eq. (\ref{EOM1}) and the term $%
\partial {\cal L}(X,\phi )/\partial \phi $ are resumed as the effective
potential $V_{eff}$, which is the thermal correction potential and is the
function of inflaton and temperature. The detailed introduction of the
temperature $T$ in warm inflation can be found in \cite
{BereraFang,Berera1999} etc. The temperature appearing in the effective
potential is that of the radiation bath and does not fall to zero thanks to
the dissipations of the inflaton to the bath provided that the dependence of
temperature in the dissipative coefficient satisfies the condition Eq. (\ref
{c}) obtained in Sec. \ref{sec3}. Under these assumptions the equation of
motion can be finally gotten:
\begin{equation}
{\cal L}_Xc_s^{-2}\ddot{\phi}+(3H{\cal L}_X+\Gamma )\dot{\phi}+V_{eff,\phi
}(\phi ,T)=0.  \label{EOM2}
\end{equation}
For simplicity, we write $V_{eff}$ as $V$ hereinafter, and the subscript $%
\phi $ denotes a derivative. We can see that the Hubble damping term is $%
{\cal L}_X$ times larger than that in canonical inflation.

The total energy density of the multicomponent Universe is
\begin{equation}  \label{totalrho}
\rho=2XK_{X}-K(X)+V(\phi,T)+Ts,
\end{equation}
where $s$ is entropy density. Through the thermodynamics relation $U=F+TS$,
we can get the free energy density of the warm inflationary Universe:
\begin{equation}  \label{freeE}
f=2XK_{X}-K(X)+V(\phi,T).
\end{equation}
Through the definition of entropy in
thermodynamics, we can get the expression for $s$:
\begin{equation}  \label{entropy}
s=-\partial f/\partial
T=-V_T(\phi,T).
\end{equation}
So, the total pressure of the Universe is
\begin{equation}  \label{totalp}
p=K(X)-V(\phi,T).
\end{equation}
Combining the total energy conservation equation $\dot\rho+3H(\rho+p)=0$ and
Eq. (\ref{EOM2}) we can get the entropy production equation:
\begin{equation}  \label{entropy1}
T\dot{s}+3HTs=\Gamma\dot{\phi}^{2}.
\end{equation}
If the thermal corrections to the potential are little enough (i.e. $b\ll1$, which can be obtained in slow-roll
valid regime; see next section), the radiation energy can be written as $\rho_r=3Ts/4$ and Eq. (%
\ref{entropy1}) is equivalent to the radiation energy density producing
equation:
\begin{equation}
\dot{\rho}_r+4H\rho _r=\Gamma \dot{\phi}^2.
\end{equation}

To get a successful inflation that has enough number of $e$-folds, we should
make
\begin{equation}
\epsilon _{_H}=-\frac{\dot{H}}{H^2}=\frac 32\frac{2XK_X+Ts}{2XK_X-K+V+Ts}\ll
1,
\end{equation}
which means
\begin{equation}
Ts\ll V,\quad XK_X\sim K\ll V
\end{equation}
i.e. the noncanonical warm inflation should be potential dominated. The
number of $e$-folds is
\begin{equation}
N=\int Hdt=\int \frac H{\dot{\phi}}d\phi \simeq -\frac 1{M_p^2}\int_{\phi
_{*}}^{\phi _{end}}\frac{V({\cal L}_X+r)}{V_\phi }d\phi ,
\end{equation}
where $r=\Gamma /3H$ is the parameter that describes the damping strength of
warm inflation.

\section{Stability analysis}

\label{sec3}

In order to make a systematic stability analysis, we define some potential
slow-roll parameters which are different from but have relations with
the Hubble slow-roll parameters \cite{Liddle1994},
\begin{equation}
\epsilon =\frac{M_p^2}{2}\left(\frac{V_{\phi}}{V}\right) ^2,~~\eta =M_p^2%
\frac {V_{\phi \phi}}{V},~~ \beta =M_p^2\frac{V_{\phi}\Gamma_{\phi}}{V\Gamma}%
,
\end{equation}
and two parameters about the temperature dependence:
\begin{equation}
b=\frac {TV_{\phi T}}{V_{\phi}},~~~ c=\frac{T\Gamma_T}{\Gamma}.
\end{equation}

We define $u=\dot{\phi}$, and Eqs. (\ref{EOM2}) and (\ref{entropy1}) can
be rewritten as
\begin{equation}  \label{EOM3}
\dot u=-{\cal L}_{X}^{-1}c_{s}^{2}\left[(3H{\cal L}_{X}+\Gamma)u+V_{\phi}(%
\phi,T)\right],
\end{equation}
and
\begin{equation}  \label{entropy2}
\dot{s}=-3Hs+\frac{\Gamma u^2}{T}.
\end{equation}
The Friedmann equation is $H^2=\frac{\rho}{3M_p^2}$.

Inflation is often associated with slow-roll approximation, which consists
of neglecting the highest order terms in Eqs. (\ref{EOM2}) and (\ref
{entropy1}). The slow-roll approximation implies that the energy is
potential dominated, the evolution of inflaton is slow and the production of
radiation is quasistatic.

We use $u_0$ , $\phi_0$ and $s_0$ to denote the slow-roll solutions that satisfy
slow-roll equations below:
\begin{equation}  \label{EOM4}
(3H{\cal L}_{X}+\Gamma)u_0+V_{\phi}(\phi,T)=0,
\end{equation}
\begin{equation}  \label{entropy3}
3H_0T_0s_0-\Gamma u_0^2=0.
\end{equation}
The variables $u$, $\phi$ and $s$ can be expanded around the slow-roll
solutions: $u=u_0+\delta u,$ $\phi=\phi_0+\delta\phi$, $s=s_0+\delta s$. The
perturbation terms $\delta u,$ $\delta s$, and $\delta \phi$ are much
smaller than the background ones $u_0$, $s_0$ and $\phi_0$. The stability is
done around the slow-roll solutions, for we should obtain the conditions to
guarantee they can really act as formal attractor solutions for the
dynamical system.

Using the new variable, $X=\frac12u^2$, then $\delta X= u\delta u$, and $%
\delta{\cal L}_{X}={\cal L}_{XX}u\delta u$. Varying the Friedmann equation
we obtain $2H_0\delta H=\frac1{3M_P^2}[{\cal L}_{X}c_{s}^{-2}u_0\delta
u+V_{\phi}\delta \phi +T_0\delta s]$. Through the thermal relation $s=-V_T$,
we have $\delta s=-V_{TT}\delta T-V_{\phi T}\delta \phi$. Then we can get
the variations of $V$, $\Gamma$ etc. by using the definition of the slow
roll parameters.

Taking the variation of Eqs. (\ref{EOM3}) (\ref{entropy2}), we can
get
\begin{equation}
\left(
\begin{array}{c}
\delta \dot{\phi} \\
\delta \dot{u} \\
\delta \dot{s}
\end{array}
\right) =E\cdot \left(
\begin{array}{c}
\delta \phi  \\
\delta u \\
\delta s
\end{array}
\right) -F.
\end{equation}
The matrices $E$ and $F$ can be expressed as
\begin{equation}
E=\left(
\begin{array}{ccc}
0 & 1 & 0 \\
A & \lambda _1 & B \\
C & D & \lambda _2
\end{array}
\right) ,~~~F=\left(
\begin{array}{c}
0 \\
\dot{u}_0 \\
\dot{s}_0
\end{array}
\right) .
\end{equation}
The matrix elements of $E$ can be calculated out:
\begin{eqnarray}
A &=&\frac{3H_0^2}{{\cal L}_X}c_s^{-2}\left[ \frac{{\cal L}_X}{{\cal L}_X+r}%
\epsilon -\eta +\frac r{{\cal L}_X+r}\beta \right.   \nonumber  \label{A} \\
&&\left. -\frac{({\cal L}_X+r)^2}rb^2+\left( {\cal L}_X+r\right) bc\right] ,
\end{eqnarray}
\begin{equation}
B=\frac{H_0T_0}{u_0}c_s^2\left[ -\frac \epsilon {\left( {\cal L}_X+r\right)
^2}-\frac c{{\cal L}_X}+\frac{{\cal L}_X+r}{{\cal L}_Xr}b\right] ,  \label{B}
\end{equation}
\begin{equation}
C=\frac{3H_0^2u_0}{T_0}\left[ \frac r{{\cal L}_X+r}\epsilon -\frac r{{\cal L}%
_X+r}\beta +\left( {\cal L}_X+r\right) (1-c)b\right] ,  \label{C}
\end{equation}
\begin{equation}
D=\frac{H_0u_0}{T_0}\left[ 6r-\frac{r{\cal L}_Xc_s^{-2}}{\left( {\cal L}%
_X+r\right) ^2}\epsilon \right] ,  \label{D}
\end{equation}
\begin{equation}
\lambda _1=-3H_0\left( 1+\frac{rc_s^2}{{\cal L}_X}\right) -H_0\epsilon \frac{%
{\cal L}_X}{\left( {\cal L}_X+r\right) ^2},  \label{lambda1}
\end{equation}
\begin{equation}
\lambda _2=-H_0(4-c)-H_0\frac{r\epsilon }{\left( {\cal L}_X+r\right) ^2}.
\label{lambda2}
\end{equation}

The slow-roll solution can be an attractor for warm inflationary dynamic
system only when the eigenvalues of the matrix $E$ are negative or possibly
positive but of order ${\cal O}(\frac \epsilon {{\cal L}_X+r})$ (i.e. we
would have slow growth) and the "forcing term" $F$ is small enough, i.e.
$|\frac{\dot{u}_0}{H_0u_0}|$ , $|\frac{\dot{s}_0}{H_0s_0}|\ll 1$. Now we
study the forcing term $F$ first. Taking the time derivative of the slow
roll equations (\ref{EOM4}) (\ref{entropy3}), we get
\begin{eqnarray}
\frac{\dot{u}_0}{H_0u_0} &=&\frac{c_s^2}\Delta \left[ \frac 1{{\cal L}_X+r}%
\left( 4-c-\frac{cr}{{\cal L}_X}\right) \epsilon +\frac 1{{\cal L}_X+r}\frac{%
4r}{{\cal L}_X}\beta \right.   \nonumber \\
&+&\left. \frac{c-4}{{\cal L}_X}\eta +3c\frac{{\cal L}_X+r}{{\cal L}_X}%
b+(c-4)\frac{\left( {\cal L}_X+r\right) ^2}{r{\cal L}_X}b^2\right] , \nonumber \\
\end{eqnarray}
\begin{eqnarray}
\frac{\dot{s}_0}{H_0s_0} &=&\frac{c_s^2}\Delta \left\{ \frac 1{{\cal L}_X+r}%
\left( 6+\frac 3{c_s^2}+\frac{3r}{{\cal L}_X}\right) \epsilon -\frac 6{{\cal %
L}_X}\eta \right.   \nonumber \\
&+&\left. \frac 1{{\cal L}_X+r}\left( \frac{9r}{{\cal L}_X}+\frac 3{c_s^2}%
\right) \beta +6\frac{\left( {\cal L}_X+r\right) ^2}{{\cal L}_Xr}b^2\right.
\nonumber \\
&+&\left. \frac{({\cal L}_X+r)}r\left[ \frac{3r(3c-1)}{{\cal L}_X}-\frac{%
3(c-1)}{c_s^2}\right] b\right\} ,\nonumber \\
\end{eqnarray}
where $\Delta \simeq (4-c)+(c+4)\frac{rc_s^2}{{\cal L}_X}$. The Hubble
parameter should also be slowly varying, i.e. $\frac{\dot{H}_0}{H_0^2}\simeq
-\frac 1{{\cal L}_X+r}\epsilon \ll 1$. Then we can get the sufficient
conditions to satisfy the above requirements:
\begin{equation}
\epsilon \ll \frac{{\cal L}_X+r}{c_s^2},~~\beta \ll \frac{{\cal L}_X+r}{c_s^2%
},~~\eta \ll \frac{{\cal L}_X}{c_s^2},~~b\ll \frac{min\{{\cal L}_X,r\}}{%
\left( {\cal L}_X+r\right) c_s^2};  \label{SR1}
\end{equation}
where $c_s^2$ is not far less than unity, and when $c_s^2\ll 1$,
\begin{equation}
\epsilon \ll \frac{{\cal L}_X+r}9,~~\beta \ll \frac{{\cal L}_X+r}9,~~\eta
\ll \frac{{\cal L}_X}{c_s^2},~~b\ll \frac r{9\left( {\cal L}_X+r\right) }.
\label{SR2}
\end{equation}
We can reach the conclusion that the slow-roll conditions in our new case
are much broader than the canonical warm inflation, let alone standard
inflation. The good features are guaranteed by the two large overdamped
terms: the larger Hubble damped term and the thermal damped term in Eq. (\ref
{EOM2}). Thus the potential can have a much broader choice and many new
models can be embedded into the cosmological inflation. This is the synergy of
the two kind effect. And noncanonical effect and thermal effect also has
competitive effect. If thermal dissipation dominates over Hubble damping
effect, i.e. $r>{\cal L}_X$, the case approximates to the canonical warm
inflationary one. In the opposite case $r<{\cal L}_X$, thermal effect is
weak but still different from cold noncanonical inflation, and the reason
we will see later in this paper. The slow-roll condition for $b$ implies
thermal correction to the inflaton potential should be small as in canonical
warm inflation \cite{Ian2008,Campo2010}. Thus the total energy density can
have a nearly separable form $\rho \simeq \rho (\phi ,X)+\rho _r$.

Now we study the matrix $E$ to give an additional slow-roll condition. Through
the slow-roll conditions we have got, we obtain that
\begin{eqnarray}
\det (\lambda I-E) &=&\left|
\begin{array}{ccc}
\lambda  & -1 & 0 \\
-A & \lambda -\lambda _1 & -B \\
-C & -D & \lambda -\lambda _2
\end{array}
\right|   \nonumber  \label{charEq} \\
&=&(\lambda -\lambda _2)\left[ \lambda (\lambda -\lambda _1)-A\right]
-BD\lambda -BC  \nonumber \\
&=&0
\end{eqnarray}
has a very small eigenvalue $\lambda \simeq \frac{BC-A\lambda _2}{\lambda
_1\lambda _2-BD-A}\ll \lambda _1,\lambda _2.$ The other two eigenvalues
satisfy $\lambda ^2-(\lambda _1+\lambda _2)\lambda +\lambda _1\lambda _2-BD=0
$. The two eigenvalues are both negative when $\lambda _1+\lambda _2<0$ and $%
\lambda _1\lambda _2-BD>0$. Finally we get
\begin{equation}
|c|<4.  \label{c}
\end{equation}
The radiation energy density is subdominated during the slow-roll
inflationary epoch: $\frac{\rho _r}V=\frac{r\epsilon }{2({\cal L}_X+r)^2}\ll
1$ which is consistent with the requirement that the inflation is potential
dominated.

\section{Cosmological perturbations}

\label{sec4}

Now we develop the theory of cosmological perturbations in the warm
noncanonical inflationary theory. The origin of density fluctuations is
thermal fluctuations and both entropy and curvature perturbations must be
present in warm inflationary scenarios. Since the energy density of
radiation is subdominant, and its fluctuation only contributes to entropy
perturbations and entropy perturbations decay on large scales \cite
{Ian2008,Cai2011,Lisa2004}, we only focus on the pure curvature
perturbation that can survive on large scales. Considering the small
perturbations, we expand the inflaton field as $\Phi (x,t)=\phi (t)+\delta
\phi (x,t)$, where $\delta \phi (x,t)$ is the linear response due to the
thermal stochastic noise $\xi $ in thermal system. In the high temperature
limit $T\rightarrow \infty $, the noise source is Markovian: $\langle \xi (%
{\bf k},t)\xi (-{\bf k^{\prime }},t^{\prime })\rangle =2\Gamma Ta^{-3}(2\pi
)^3\delta ^3({\bf k}-{\bf k^{\prime }})\delta (t-t^{\prime })$ \cite
{Lisa2004,Gleiser1994}. Introducing the noise term and substituting the
expansion of inflaton we can get a second order Langevin equation:
\begin{eqnarray}
{\cal L}_Xc_s^{-2}\left[ \ddot{\phi}(t)+\delta \ddot{\phi}({\bf x}%
,t)\right] +\left( 3H{\cal L}_X+\Gamma \right) \left[ \dot{\phi}(t)+\delta
\dot{\phi}({\bf x},t)\right]   \nonumber \\
+V_\phi +\left( V_{\phi \phi }-{\cal L}_X\frac{\nabla ^2}{a^2}\right) \delta
\phi ({\bf x},t) =\xi ({\bf x},t). \nonumber \\
\end{eqnarray}
Then we take the Fourier transform and obtain the evolution equation for the
fluctuation:
\begin{equation}
{\cal L}_Xc_s^{-2}\ddot{\delta \phi _{{\bf k}}}+\left( 3H{\cal L}_X+\Gamma
\right) \delta \dot{\phi _{{\bf k}}}+\left( {\cal L}_X\frac{k^2}{a^2}%
+m^2\right) \delta \phi _{{\bf k}}=\xi _{{\bf k}}.  \label{pEOM2}
\end{equation}
The second order Langevin equation is hard to solve and we only want to get
the power spectrum when horizon crossing. Horizon crossing is well inside
the slow-roll inflationary regime \cite{LiddleLyth} and the slow-roll regime is
overdamped so the inertia term can be neglected. Then the Langevin equation (%
\ref{pEOM2}) can be reduced to first order as in \cite
{Berera2000,TalyorBerera}:
\begin{equation}
(3H{\cal L}_X+\Gamma )\delta \dot{\phi _{{\bf k}}}+\left( {\cal L}_X\frac{k^2%
}{a^2}+m^2\right) \delta \phi _{{\bf k}}=\xi _{{\bf k}}.  \label{pEOM3}
\end{equation}
The approximate analytic solution is
\begin{eqnarray}
\delta \phi _{{\bf k}}(t) &\approx &\left\{ \frac 1{3H{\cal L}_X+\Gamma }%
\int_{t_0}^t\exp \left[ \frac{t^{\prime }-t_0}{\tau (\phi _0)}\right] \xi (%
{\bf k},t^{\prime })dt^{\prime }\right.   \nonumber \\
&&\left. +\delta \phi ({\bf k},t_0)\right\} \exp \left[ -\frac{t-t_0}{\tau
(\phi _0)}\right]
\label{phik}
\end{eqnarray}
where $\tau (\phi )=\frac{3H{\cal L}_X+\Gamma }{{\cal L}_Xk^2/a^2+m^2}$,
which describes the efficiency of the thermalizing process. The relation
between physical wave number $k_p$ and comoving wave number $k$ is $k_p=k/a$.
In the expanding Universe, we can see from Eq. (\ref{phik}) that the larger $%
k_p^2$ is, the faster the relaxation rate is. If $k_p^2$ is sufficiently
large for the mode to relax within a Hubble time, then that mode
thermalizes. As soon as the physical wave number of a $\delta \phi (x,t)$
field mode becomes less than $k_F$ , it essentially feels no effect of the
thermal noise $\xi ({\bf k},t)$ during a Hubble time \cite{Berera2000}.
Based on the criterion, the freeze-out physical momentum $k_F$ is defined
as $\frac{{\cal L}_X{\bf k}_F^2+m^2}{(3H{\cal L}_X+\Gamma )H}=1$. The mass
term is negligible compared to other terms in slow-roll inflation. Then we
can work out
\begin{equation}
k_F=\sqrt{\frac{3H^2({\cal L}_X+r)}{{\cal L}_X}}.  \label{kf1}
\end{equation}
Based on the field perturbation relation $\delta \phi ^2=\frac{k_FT}{2\pi ^2%
}$ in warm inflation \cite{Berera2000,TalyorBerera}, and using $P_R=\left(
\frac H{\dot{\phi}}\right) ^2\delta \phi ^2$, we can finally get the scalar
power spectrum in warm noncanonical inflationary model:
\begin{equation}
P_R=\frac{H^3T}{2\pi ^2u^2}\sqrt{\frac{3({\cal L}_X+r)}{{\cal L}_X}}=\frac{%
9H^5T({\cal L}_X+r)^{\frac 52}}{2\pi ^2V_\phi ^2}\sqrt{\frac 3{{\cal L}_X}}.
\label{PR}
\end{equation}
CMB observations provide a good normalization of the scalar power spectrum $%
P_R\approx 10^{-9}$ on large scales, so we can see from the $({\cal L}%
_X+r)^{5/2}$ in the numerator that the energy scale when horizon crossing
can be depressed by both the noncanonical effect and thermal effect, which
is good news to the assumption that the Universe inflation can be
described well by effective field theory. The spectral index
\begin{equation}
n_s-1=\frac{d\ln P_R}{d\ln k}
\end{equation}
is given by
\begin{widetext}
\begin{equation}
n_s-1=\alpha _1\frac{c_s^2}{{\cal L}_X+r}\epsilon +\alpha _2\frac{c_s^2}{%
{\cal L}_X}\eta +\alpha _3\frac{c_s^2}{{\cal L}_X+r}\beta +\alpha _4\frac{%
c_s^2({\cal L}_X+r)}{min\{{\cal L}_X,r\}}b+\alpha _5\frac{c_s^2({\cal L}%
_X+r)^2}{{\cal L}_Xr}b^2,
\end{equation}
where the expressions for $\frac{\dot{T}}{HT}$, $\frac{\dot{{\cal L}_X}}{H%
{\cal L}_X}$ and $\frac{\dot{r}}{H({\cal L}_X+r)}$ are used. The parameters $%
\alpha _1$, $\alpha _2$, $\alpha _3$, $\alpha _4$ and $\alpha _5$ are given
by:
\begin{equation}
\alpha _1=-3c_s^2+\frac{rc_s^2}{2({\cal L}_X+r)}+\frac 1\Delta \left\{
\left( 6+\frac 3{c_s^2}+\frac{3r}{{\cal L}_X}\right) \left[ 1+\frac{cr}{%
2\left( {\cal L}_X+r\right) }\right] -3\left( 4-c-\frac{cr}{{\cal L}_X}%
\right) \left[ 2+\frac{r\left( c_s^{-2}-1\right) }{2\left( {\cal L}%
_X+r\right) }\right] \right\} ,
\end{equation}
\begin{equation}
\alpha _2=\frac 1\Delta \left\{ -6-\frac{3cr}{{\cal L}_X+r}+\left(
12-3c\right) \left[ 2+\frac r{2\left( {\cal L}_X+r\right) }%
(c_s^{-2}-1)\right] \right\} ,
\end{equation}
\begin{equation}
\alpha _3=\frac 1\Delta \left\{ \left( \frac{9r}{{\cal L}_X}+\frac 3{c_s^2}%
\right) \left[ 1+\frac{cr}{2\left( {\cal L}_X+r\right) }\right] \right.
\left. -12\frac r{{\cal L}_X}\left[ 2-\frac r{2\left( {\cal L}_X+r\right) }%
(c_s^{-2}-1)\right] -\frac{4r}{{\cal L}_X}\right\} ,
\end{equation}
\begin{eqnarray}
\alpha _4 &=&\frac 1\Delta \frac{min\{{\cal L}_X,r\}}r\left\{ \frac{%
3(2rc-6r-3)}{{\cal L}_X}-\frac{cr}{2\left( {\cal L}_X+r\right) }\left[ \frac{%
3r(3c-1)}{{\cal L}_X}-\frac{3(c-1)}{c_s^2}\right] \right\}   \nonumber \\
&&-\frac 1\Delta \frac{min\{{\cal L}_X,r\}}{{\cal L}_X}\left[ 2+\frac r{2(%
{\cal L}_X+r)}(c_s^{-2}-1)\right] -\frac 12c\frac{min\{{\cal L}_X,r\}}{c_s^2(%
{\cal L}_X+r)},
\end{eqnarray}
\begin{equation}
\alpha _5=\frac 3\Delta \left\{ 2\left[ 1+\frac{cr}{2({\cal L}_X+r)}\right]
-(c-4)\left[ 2+\frac{r(c_s^{-2}-1)}{2({\cal L}_X+r)}\right] \right\} .
\label{alpha5}
\end{equation}
\end{widetext}
The five parameters above are all of order unity, so we can find that $n_s-1$
is of order ${\cal O}\left( \frac{\epsilon c_s^2}{{\cal L}_X+r}\right) \ll 1$%
, where $\epsilon $ refer to the slow-roll parameters in general. We
obtained a nearly scale-invariant power spectrum that is consistent with
observations. The running of the spectral $\alpha _s=\frac{dn_s}{d\ln k}$
is calculated to find that it is of order $\left( \frac{\epsilon c_s^2}{%
{\cal L}_X+r}\right) ^2\ll (n_s-1)$, which coincides with observations
qualitatively. And we can study some concrete models in the new theory
numerically and fix the physical quantities by comparing with new
observations given by PLANCK satellite in the future.

The tensor perturbations do not couple to the thermal background, and so
gravitational waves are only generated by the quantum fluctuations as in
standard inflation \cite{TalyorBerera}
\begin{equation}
P_T=\frac 2{M_p^2}\left( \frac H{2\pi }\right) ^2.
\label{tensorperturbation}
\end{equation}
The spectral index of tensor perturbation is $\label{ng}n_T=-2\frac \epsilon
{1+r}$, and the tensor-to-scalar ratio is
\begin{equation}
R=\frac{P_T}{P_R}=\frac HT\frac{2\epsilon \sqrt{{\cal L}_X}}{\sqrt{3}\left(
{\cal L}_X+r\right) ^{5/2}}.  \label{ratio}
\end{equation}
We can see that the tensor perturbation can be much weaker thanks to both
the noncanonical effect and thermal effect if both the effects are strong,
which is another synergy of both effects. Considering the slow-roll
condition $\epsilon <\frac{{\cal L}_X+r}{c_s^2}$, we can get the upper bound
of the tensor-to-scalar ratio $R<\frac HT\frac{2({\cal L}_X+2X{\cal L}_{XX})%
}{\sqrt{3{\cal L}_x}\left( {\cal L}_X+r\right) ^{3/2}}$. As BICEP2
suggests recently, the tensor-to-scalar ratio is significant at a level $%
R=0.2\pm 0.05$ \cite{BICEP2}, the upper bound of R in our case should be
large enough. We can obtain that a significant R prefers a weaker ($r\ll 1$)
noncanonical warm inflationary scenario with a big sound speed ($c_s$ is
order of unity). The insignificant non-Gaussianity suggested by PLANCK \cite
{PLANCK2} also prefers a big sound speed of the noncanonical inflaton \cite
{Cai2011,Franche2010}. The amount of expansion is $\Delta N\simeq 4$ while
the scales corresponding to $2\leq l\leq 100$ are leaving the horizon, the
corresponding variation of field is $\frac{\Delta \phi }{M_p}=\frac{\dot{\phi%
}\Delta N}{M_pH}\simeq 5.2(\frac TH)^{1/2}(1+\frac r{{\cal L}_X}%
)^{1/4}R^{1/2}$. The field variation can be smaller than Planck scale
opposite to standard inflation [$\frac{\Delta \phi }{M_p}=0.5(\frac R{0.1}%
)^{1/2}$ \cite{LiddleLyth}] in strong regime of warm inflation, which can
cure the overlarge amplitude of inflaton in standard inflation. The
consistency equation becomes $R=-\frac HT\frac{{{\cal L}_X}^{1/2}(1+r)}{%
\sqrt{3}({\cal L}_X+r)^{5/2}}n_T$, which is not a fixed relation as in
standard inflation ($R=-6.2n_T$) \cite{LiddleLyth} anymore.

The radiation energy density and the universal temperature has the
Stefan-Boltzmann relationship $\rho_r=\pi^2g_{\ast}T^4/30$, and using the
slow-roll equations we obtained
\begin{equation}  \label{TH}
\frac{T}{H}=\left(\frac{r}{g_{\ast}P_R}\right)^{1/3}\left(\frac{45}{4\pi^2}%
\right)^{1/3}\left[3\left(1+\frac r{{\cal L}_X}\right) \right]^{1/6}
\end{equation}
from the scalar power spectrum. The ratio $T/H$ is smaller than that of warm
canonical inflation \cite{Ian2008,ZhangZhu} for the variable ${\cal L}_X$ in
the denominator in the last factor. We can see that a larger $r/{\cal L}_X$
can enhance the ratio $T/H$; thus, the thermal effect is more obvious and the
case is opposite when we have a smaller $r/{\cal L}_X$, which is the
competitive effect of the noncanonical effect and thermal effect. The
criterion for the happening of warm inflation $T>H$ can be easily and
sufficiently satisfied by $r>g_{\ast}P_R$ by analyzing Eq. (\ref{TH}).
Considering that $g_{\ast}$ is of order ${\cal O}(10^2)$ and $P_R$ is of
order ${\cal O}(10^{-9})$, we can find very small amounts of dissipation that can
result in warm inflation. So the warm inflation can describe the very early
Universe more realizably and even in weak dissipative regime $r\ll{\cal L}_X$%
, and the thermal fluctuation amplitude dominates over its quantum counterpart,
which is the consequence of Eqs. (\ref{PR}) and (\ref{TH}).

\section{Conclusions}

\label{sec5} We summarize with a few remarks. We develop a theory of warm
noncanonical inflationary scenario and generalize the scope of the inflation.
Through the action of the warm Universe system, we get the equation of
motion for the inflaton and other basic equations of the new scenario. The
Hubble damping term is enhanced by an important physical quantity ${\cal L}_X
$ in noncanonical field. The stability analysis is made to give out broader slow-roll conditions thanks to the thermal and noncanonical effect.
We obtain a new form but still nearly scale-invariant scalar power spectrum
and we find the energy scale during horizon crossing can be depressed by the
synergy of the two effects. The tensor-to-scalar ratio can be significant in
weak noncanonical warm inflation with a big sound speed and insignificant
in the opposite case. Warm noncanonical inflation in strong regime is also
a kind of scenario to cure the eta problem and overlarge amplitude of the
inflaton. We will focus on some concrete models of the new theory to give
more precise comparison with the observations in the future. And the
detailed issue of non-Gaussianity in the new scenario also deserves more
cognition and research.

\acknowledgments This work was supported by the National Natural Science Foundation of China (Grants No. 11175019 and No. 11235003).

\end{document}